# Chiral-induced unidirectional spin-to-charge conversion

## Authors

Ashish Moharana[1], Yael Kapon[2], Fabian Kammerbauer[1], David Anthofer[1], Shira Yochelis[2], Mathias Kläui[1], Yossi Paltiel[2], and Angela Wittmann[1*]

## Affiliations

[1]Institute of Physics, Johannes Gutenberg University Mainz, Mainz 55128, Germany

[2]Institute of Applied Physics and Center for Nanoscience and Nanotechnology, The Hebrew University of Jerusalem, Jerusalem 9190401, Israel

*Corresponding author: a.wittmann@uni-mainz.de

## Abstract

The observation of spin-dependent transmission of electrons through chiral molecules has led to the discovery of chiral-induced spin selectivity (CISS). The remarkably high efficiency of the spin polarizing effect has recently gained significant interest due to the high potential for novel sustainable hybrid chiral molecule magnetic applications. However, the fundamental mechanisms underlying the chiral-induced phenomena remain to be understood fully. In this work, we explore the impact of chirality on spin angular momentum in hybrid metal/ chiral molecule thin film heterostructures. For this, we inject a pure spin current via spin pumping and investigate the spin-to-charge conversion at the hybrid chiral interface. Notably, we observe a chiral-induced unidirectionality in the conversion. Furthermore, angle-dependent measurements reveal that the spin selectivity is maximum when the spin angular momentum is aligned with the molecular chiral axis. Our findings validate the central role of spin angular momentum for the CISS effect, paving the path toward three-dimensional functionalization of hybrid molecule-metal devices via chirality.

## Introduction

The spin-dependent charge transport through chiral molecules gives rise to a well-established phenomenon known as chiral-induced spin selectivity (CISS). Over the past decade, studies have consistently demonstrated chiral-induced spin polarization with efficiencies of up to more than 70% (1-6). Extensive investigations into the CISS effect have so far primarily focused on photoemission spectroscopy and transport measurements. The initial observation of spin-polarized photoelectron emission through a polyalanine layer revealed a dependence on the chirality of the peptide molecules (7). Another approach to exploring the CISS effect has involved local transport measurements (8-11).

In addition to the CISS effect in the transport of charge carriers, it is crucial to note that chiral molecules can also impact the properties of underlying metal thin films significantly. The hybrid interface between molecules and metal thin films has been a fruitful playground for engineering interfacial properties through molecular design (12,13). Hybridization between molecules and metal thin films leads to changes in the electronic and magnetic properties at the hybrid interface (14-19). Along with the widely studied effects of charge transfer and exchange coupling at hybrid interfaces, hybridization also impacts the effective spin-orbit coupling (SOC) in the hybrid system. A prominent example of the CISS effect upon hybridization is the manipulation of magnetization when chiral molecules are adsorbed on metallic and ferromagnetic surfaces (11,17).

Numerous experimental and theoretical efforts have tried to understand the underlying microscopic origins of the CISS effect. However, a fundamental debate persists in elucidating its microscopic mechanisms. Studies have demonstrated that SOC originating from the helical structure of chiral molecules plays a pivotal role in the CISS effect (18,19). However, the magnitude of the SOC within these molecules is often too small to account for the significant spin-filtering effect observed in experiments. To address this challenge, several studies have focused on the role of SOC at the hybrid interface between heavy metals and chiral molecules. These studies have shown that CISS is a result of the interaction between the high SOC of the metal electrode and charge distribution at the interface of the molecule and metal (20-26). Several reports have attempted to provide a theoretical description of chiral structures by incorporating electron-phonon and electron-electron interactions. These interactions are a result of the SOC, leading to the exchange splitting of the spin channels within the structure, ultimately contributing to the phenomenon of spin filtering (27,28). On the other hand, recent studies have shown that the CISS effect can result from the chiral-induced orbital polarization effect of chiral molecules. In this framework, the topological electronic property of chiral molecules is characterized by the locking of spin and orbital angular momentum. Electrons with orbital angular momentum compatible with the molecular chirality find it easier to enter the chiral layer. This orbital polarization effect induces spin polarization mediated by the SOC in the heavy metal contact resulting in spin selectivity in the hybrid chiral molecule metal system (29-31).

Thus, conclusive evidence has been lacking in establishing whether helical chiral molecules effectively function as spin or orbital polarizers. To experimentally verify the effect of chiral molecules on spin angular momentum, in this study, we investigate the CISS effect at a hybrid chiral interface using a pure spin current. For this, we inject a pure spin current from a ferromagnetic insulator into a hybrid metal/ chiral molecule bilayer at ferromagnetic resonance. The SOC of the hybrid layer converts the spin current into an electromotive force via the inverse spin Hall effect (ISHE), measurable as a voltage signal across the metal layer (32). The results show a clear signature of the CISS effect confirming that spin angular momentum plays a significant role in the CISS effect.

## Results and Discussion

In this study, we perform spin pumping experiments in the ferromagnetic insulator ($Y_3Fe_5O_{12}$, YIG, 103 nm)/ heavy metal (Au, 4 nm) bilayer structures to generate a pure spin current in a non-magnetic metal layer. The samples are placed on a grounded coplanar waveguide to excite the precession of the magnetization with an AC microwave magnetic field (10 GHz). At ferromagnetic resonance, spin angular momentum is injected from the ferromagnetic layer into the adjacent non-magnetic layer (33). Due to the inverse spin Hall effect, the resulting pure spin current generates an electromotive force perpendicular to the spin polarization and the spin current direction that can be detected as a voltage signal $V_{ISHE}$. The direction of the spin polarization in the pure spin current can be controlled by changing the orientation of the magnetization of the ferromagnet using an external magnetic field. A schematic illustration of the geometry of the spin pumping measurement is shown in Fig. 1**A**. Figure 1**B** shows the characteristic Lorentzian resonance shape of the inverse spin Hall effect voltage $V_{ISHE}$ signal around the ferromagnetic resonance as a function of the applied magnetic field. Here, the magnetic field was applied at an angle of 60° (+B) and 240° (−B) (Fig. 1**B**).

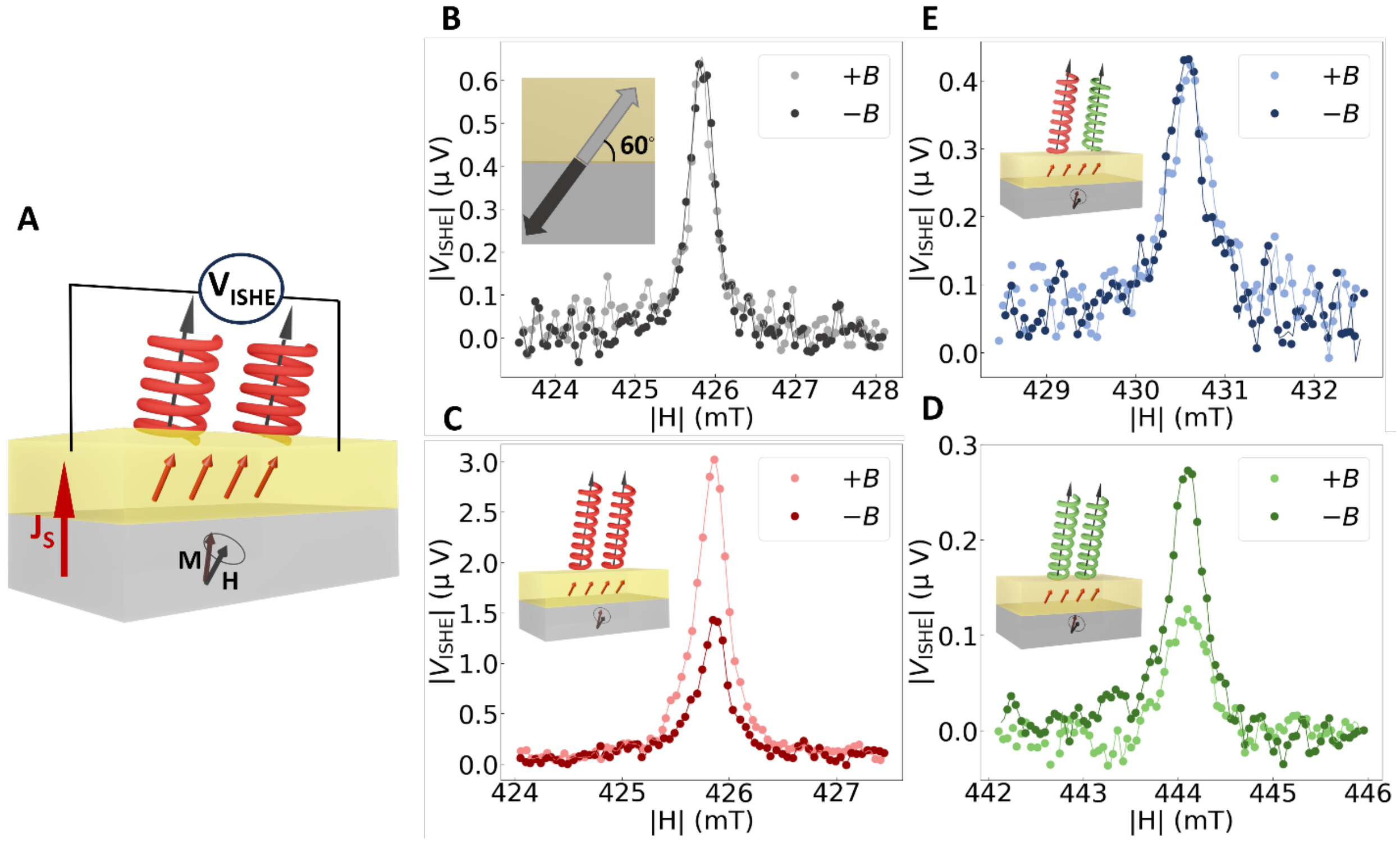

**Fig. 1. Inverse Spin Hall Effect (ISHE) in Au/YIG bilayer: (A)** Schematic representation of spin pumping at the hybrid interface between Au and chiral molecules and detection of pure spin current through the Inverse Spin Hall Effect (ISHE) by measuring the DC voltage across the Au layer. **(B)** ISHE voltage measurement through the spin pumping along the spin polarization at an angle of $\alpha$ = 60°. The absolute magnitude of ISHE for both positive and negative magnetic fields without molecules. **(C)** Plot illustrating the Inverse Spin Hall Effect (ISHE) for both positive and negative magnetic fields with the 36-L alpha helix polyalanine molecule. **(D)** Plot illustrating the Inverse Spin Hall Effect (ISHE) for both positive and negative magnetic fields with the 36-D alpha helix polyalanine molecule. **(E)** Plot depicting the ISHE effect for both positive and negative magnetic fields with a racemic mixture of polyalanine molecules.

There is no significant difference in the absolute magnitude of $V_{ISHE}$ at ferromagnetic resonance in YIG/Au between the positive (dark grey) and negative (light grey) external magnetic field. We note, that the voltage signal $V_{ISHE}$ reverses its sign when the polarity of the magnetic field is inverted, consistent with the symmetry of the ISHE (see supplementary information). However, to compare the absolute magnitude in the ISHE between positive and negative magnetic fields, the data is shown in absolute values.

## Hybrid chiral systems

To probe and quantify the impact of the CISS effect on the spin-to-charge conversion in the hybrid heavy-metal/ chiral molecule system, a self-assembled monolayer of 36-L alpha helix polyalanine was adsorbed on the YIG/Au sample (see schematic inset in Fig. 1**C**). Figure 1**C** shows the ISHE voltage signal of the hybrid chiral device around ferromagnetic resonance for a positive (light red) and negative (dark red) magnetic field along 60˚. In stark contrast to the bare YIG/Au device, there is a significant change in the magnitude of the ISHE voltage signal when the polarity of the magnetic field is reversed in the hybrid chiral device. To probe the role of the chirality on the observed spin selectivity, we investigated a hybrid chiral device with chiral molecules of the opposite handedness. Figure 1**D** shows also a significant difference in the ISHE voltage signal of the hybrid chiral YIG/Au/36-D alpha helix polyalanine device between ferromagnetic resonance in a positive (light green) and negative (dark green) magnetic field along 60˚. However, importantly, the sign of the difference in the magnitude of the ISHE is reversed between the two opposite chiralities of the molecules. This implies that the spin selectivity in the spin-to-charge conversion efficiency for a given spin polarization direction depends on the chirality of the hybrid chiral system. Furthermore, to disentangle the effect of the chirality of the molecules from the common chirality-independent hybridization effects of molecules on metal surfaces, we have performed a control experiment on a hybrid system with a racemic mixture of the polyalanine molecules. The racemic mixture consists of equal fractions of both optical rotations. The ISHE voltage signal for the device with the racemic mixture is shown in Fig. 1**E**. Akin to the bare YIG/Au sample, the magnitude of the ISHE voltage signal $V_{ISHE}$ at ferromagnetic resonance does not depend on the polarity of the magnetic field. Consequently, the observed asymmetry in the spin-to-charge conversion efficiency in the hybrid chiral system is directly linked to the chirality of the molecules, presenting a distinct signature of the CISS effect. The dependence of the magnitude of the ISHE voltage signal $V_{ISHE}$ on the polarity of the spin polarization in the pure spin current implies that the chirality of the molecules significantly impacts the effective spin-orbit coupling at the hybrid chiral interface. This result is consistent with the theoretical prediction of chiral charge transfer between the heavy metal-molecule interface resulting in a chiral distribution of charges in metals (34).

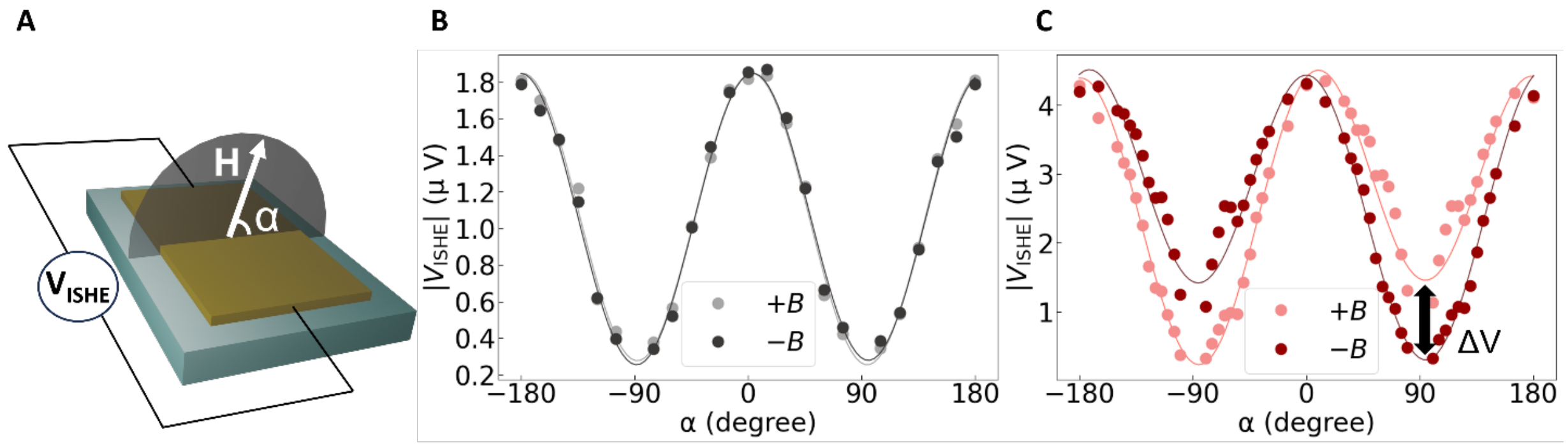

**Fig. 2. Angular Dependence of ISHE Measurements. (A)** Schematics representation of the plane of measurement where H represents the applied field, and α signifies the angle of measurement **(B)** Plot illustrating the angular dependence of the Inverse Spin Hall Effect (ISHE) amplitude measurement with pristine Au. **(C)** Plot depicting the asymmetry in $V_{ISHE}$ angular measurement with 36 alpha helix alanine on top of Au. Here ΔV represents the difference in the $V_{ISHE}$ for different spin orientations with positive and negative fields.

**Angle-dependent measurements**

In the next step, we examine the full out-of-plane angle dependence of the inverse spin Hall effect signal to investigate the symmetry of the spin selectivity in more detail. To maximize the detected voltage signal $V_{ISHE}$, we focus on the angle-dependence within the plane of rotation perpendicular to the voltage leads in the two-terminal devices as shown schematically in Fig. 2**A**. Figure 2**B** shows the magnitude of the ISHE voltage signal $V_{ISHE}$ of the bare YIG/Au device as a function of the out-of-plane angle $\alpha$ of the external magnetic field for positive and negative polarity (light and dark grey respectively). As expected from the symmetry of the ISHE, the magnitude of the ISHE $|V_{ISHE}|$ does not depend on the polarity of the external magnetic field. In contrast to this, the angle-dependent magnitude $|V_{ISHE}|$ in the hybrid homochiral device differs between the positive and negative (light and dark red respectively) external magnetic field as shown in Fig. 2**C**. Furthermore, the asymmetry is inverted for $\alpha = \pm 90^{\circ}$.

The experimental data fits well to a $\cos(2\alpha) + \sin(\alpha)$ function. From the fits (solid lines), we can extract the maximum asymmetry in the ISHE voltage $|\Delta V_{ISHE}|$. While we find $|\Delta V_{ISHE}| = 23 \pm 31$ nV for the pristine YIG/Au device, the asymmetry $|\Delta V_{ISHE}| = 1376 \pm 56$ nV is significantly larger in the hybrid homochiral device. This result further confirms the presence of a chiral-induced unidirectional component in the inverse spin Hall effect in the hybrid homochiral system. Moreover, we can quantify the effective spin selectivity ($S$),

$$S = \frac{V_{ISHE}(+B) - V_{ISHE}(-B)}{V_{ISHE}(+B) + V_{ISHE}(-B)}.$$

Figure 3**A** shows the angle-dependence of the extracted spin selectivity of the L and D rotation of the hybrid homochiral and the hybrid achiral system with the racemic mixture. The hybrid

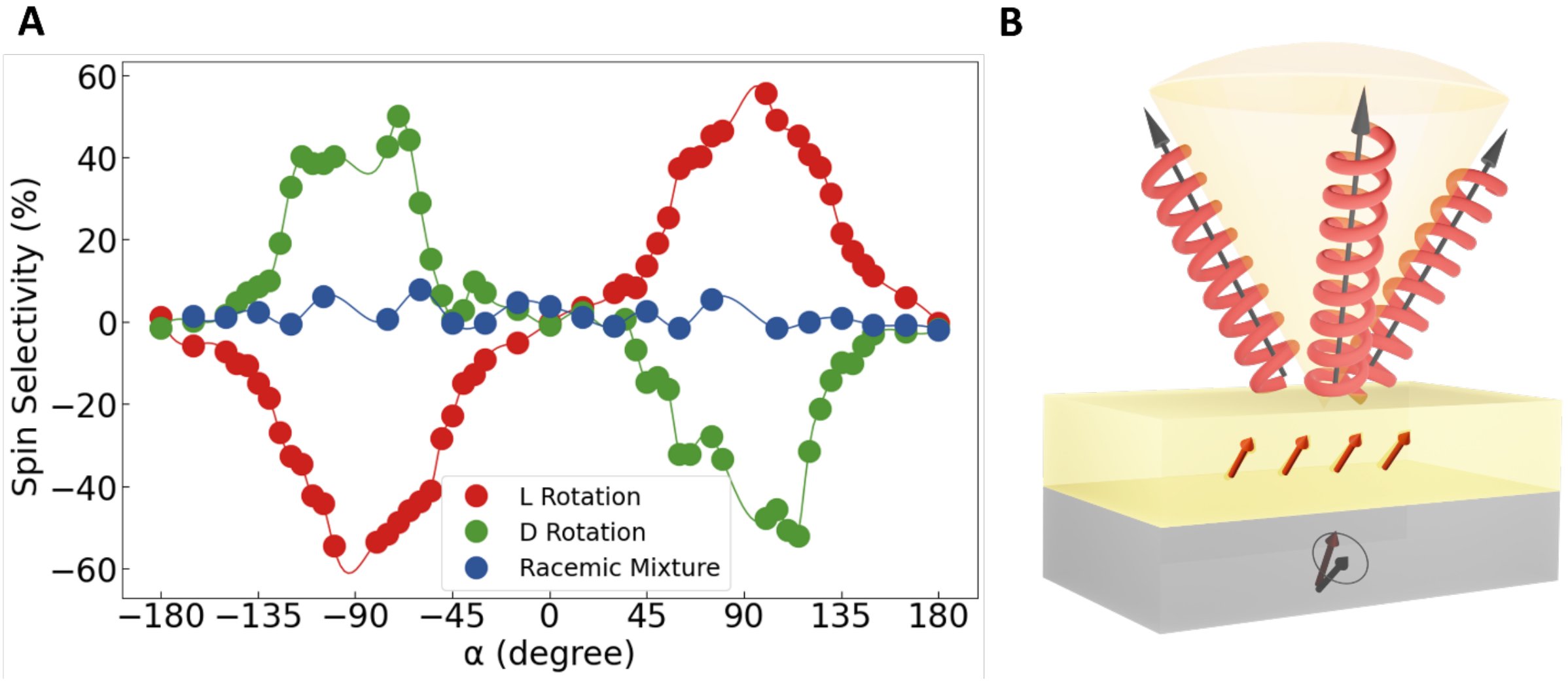

**Fig. 3. Angular Dependence of Spin selectivity Effect: (A)** Spin selectivity effect represents the change in $V_{ISHE}$ for different spin orientations for both positive and negative magnetic fields normalized for $V_{ISHE}$ for L (red) and D (green) rotation of 36 alpha helix polyalanine and the racemic mixture (blue) on Au. The solid line represents cubic spline interpolation. **(B)** Schematic depicting the conical orientation of molecules on Au, where the molecules are arranged at an angle of 60° with in-plane.

achiral system (blue) does not show any significant spin selectivity at any angle. In contrast, the L and D rotation of the hybrid homochiral systems (red and green respectively) show a sizeable spin selectivity of up to nearly 60%. The magnitude of the maximum spin selectivity agrees well with previous reports of the CISS effect in charge currents in the literature (35,36). Importantly, the sign of the spin selectivity is reversed for the two opposite helicities of the L and D molecules.

We note, that the spin selectivity sharply increases at $|\alpha| = 50^{\circ}$, has a maximum around $|\alpha| = 90^{\circ}$, and abruptly decreases again at $|\alpha| = 130^{\circ}$. In self-assembled monolayers, the polyalanine molecules do not arrange perfectly perpendicular to the sample surface but have been shown to tilt to approximately 60° with respect to the surface (22,37,38). The direction of the tilt in the different domains in the devices is uniformly distributed. Thus, we can presume that the molecules are all oriented within a cone normal to the surface as depicted schematically in Fig. 3**B**. The ISHE measurement probes the signal averaged over many domains with different azimuthal angles. As a result, the agreement between the strong and non-uniform angle dependence of the efficiency of the chiral-induced effect with the overall orientation of the molecules implies that the observed effect is vectorial. This is akin to previous reports of vectorial spin filtering via the CISS effect in charge currents (39).

In summary, we have performed spin pumping experiments to probe the impact of chirality on pure spin currents. Our findings show clear signatures of the CISS effect in the electrical detection of pure spin currents. In particular, we report chiral-induced strongly unidirectional spin-to-charge conversion via the inverse spin Hall effect. Through angle-dependent spin injection, we establish that the maximum spin selectivity efficiency occurs when the spin orientation aligns with the orientation of the molecular axis. Therefore, the chiral-induced anisotropy in the interfacial spin-orbit coupling acts vectorially along the axis of the molecules. These results imply that the CISS effect depends significantly on the electrons' spin angular momentum, opening up a pathway toward targeted vectorial manipulation of hybrid spintronic devices via chirality.

## Materials and Methods

The chiral molecules were chosen to be 36 - L/D $\alpha$-helix polyalanine (L/D-AHPA) [[H] CAAAAKAAAAKAAAAKAAAAKAAAAKAAAAKAAAAK-[OH]] molecules (C stands for cysteine, A for alanine, and K for lysine), manufactured by Sigma–Aldrich. A 1 mM solution was prepared in absolute ethanol and used in the experiments. The racemic mixture was produced by mixing equal parts of the L and D solutions.

## Monolayer adsorption

Sample cleaning was performed with boiling acetone for 10 minutes followed by boiling isopropanol for 10 minutes and subsequently double distilled water. To prepare the exposed Au surface for adsorption, it was immersed in ethanol for 20 minutes to reduce the produced oxides. Afterward, the molecules were chemically adsorbed through their thiol head group onto the surface via a 72-hour immersion, for a dense and organized monolayer, in a 1 mM solution of the molecules in absolute ethanol in a nitrogen environment. The substrates were washed in dry ethanol and dried off. XPS measurements were performed to characterize the monolayer growth (see supplementary information).

**Sample fabrication process**

In the fabrication of the YIG/Au sample utilized in this study, a 4 nm-thick layer of gold (Au) was sputter-deposited onto YIG substrates with a thickness of 103 nm, grown using the Liquid Phase Epitaxy (LPE) technique. The sputter deposition of the Au layer was carried out in an argon (Ar) plasma at a controlled rate of 0.9 Å/s with a base pressure of 5e−8 mbar. Standard photolithography and Ar ion etching were used to fabricate the 100 $\mu$m wide and 800 $\mu$m long bar structures. A metallic shadow mask was used to deposit Au electrodes. Sputtering was used to deposit 5 nm of Cr and 50 nm of Au electrodes.

**Measurement procedure**

The YIG/Au sample was mounted on top of a strip line of the grounded coplanar waveguide. The input microwave power remained consistently fixed at 10 dBm, and the microwave frequency was set to 10 GHz. The ferromagnetic resonance and ISHE measurements were carried out using a lock-in technique modulating the amplitude of the microwave signal at a modulation frequency of 1.5 kHz. We have verified that the amplitude of the microwave absorption at ferromagnetic resonance is comparable for both polarities of the magnetic field (see supplementary information).

## Acknowledgments

The XPS measurements of the AHPA monolayer on Au were taken by Dr. Vitaly Gutkin of the Hebrew University Center for Nanoscience and Nanotechnology. We thank Sumit Ghosh for helpful discussions. The authors acknowledge funding by the Carl-Zeiss-Stiftung (HYMMS P2022-03-044). AM, FK, MK, and AW thank the German Research Foundation (SFB TRR 173 Spin+X 268565370 projects A01, B02, B14). YK, SY, and YP acknowledge funding from the Ministry of Science (MOS) Israel.